\documentclass[11pt,a4paper]{article}
\usepackage[margin=2.35cm]{geometry}
\usepackage{amsmath,amssymb,bm,mathtools}
\usepackage{graphicx,booktabs,array,microtype}
\usepackage{hyperref,xcolor}
\usepackage{tikz,pgfplots}
\usepackage{authblk}
\usepackage[numbers,sort&compress]{natbib}
\usepackage{enumitem}
\usetikzlibrary{arrows.meta,positioning,calc}
\pgfplotsset{compat=1.17}
\hypersetup{colorlinks=true,linkcolor=blue,citecolor=blue,urlcolor=blue}
\title{\bfseries From Colour--Casimir Recoupling to Near-Threshold Scattering:\\
Algebraic Diagnostics and a Pion-Mass Test for the $T_{cc}^{+}$ Channel}
\author{M. Monemzadeh\thanks{Corresponding author. E-mail: \href{mailto:monem@kashanu.ac.ir}{monem@kashanu.ac.ir}}}
\author{N. Tazimi}
\affil{Department of Physics, University of Kashan, Kashan, Iran}
\date{}

\begin{document}
\maketitle

\begin{abstract}
The near-threshold $T_{cc}^{+}$ is a clean testing ground for how constituent-level colour structure connects to hadronic scattering. We report three results and state what each does and does not establish.
(i)~\emph{Colour algebra.} In the two-dimensional colour-singlet basis of $cc\bar u\bar d$, a general pairwise colour operator has direct $DD^*$ matrix element $V_{MM}=-\tfrac43(v_{13}+v_{24})$, containing only the two intra-meson coefficients, while meson--hidden-colour mixing is $V_{MH}^{\rm off}=\tfrac{\sqrt2}{3}\Delta_v$ with $\Delta_v=v_{14}+v_{23}-v_{12}-v_{34}$; uniform edges give no mixing. The projection is exact within this basis and is checked numerically on 10\,000 random edge assignments. A same-Hamiltonian decomposition shows that it re-expresses the central colour operator ($116.46$ MeV off-diagonal) rather than adding an independent dynamical input.
(ii)~\emph{Negative dynamical results.} For Coulomb-type Gaussian edges the hidden-colour diagonal lies below the meson diagonal, so the attractive second-order shift $-2\Delta_v^2/(9\Delta_H)$ does not apply. In a Gaussian constituent Hamiltonian calibrated to $M_D$ and $M_{D^*}$, the lowest compact level lies $148.6$ MeV above its model $DD^*$ threshold; this concerns the compact sector only and is not a coupled-channel pole.
(iii)~\emph{Pion-mass test.} Couplings fixed at the $m_\pi\simeq280$ MeV lattice pole ($-5.2$ MeV) are carried to $m_\pi\simeq391$ MeV by changing only the reduced mass and compact-sector gap, both refitted to the lattice $D$ and $D^*$ masses. The model keeps a shallow bound state at about $-4$ to $-4.6$ MeV, independent of the regulator, whereas the lattice reports a virtual state about $62$ MeV below threshold. Colour algebra, a contact term, and a compact gap thus cannot supply the observed light-quark-mass dependence; pion exchange and the left-hand cut are required.
A one-channel separable Lippmann--Schwinger benchmark ($\Lambda=650$ MeV, $\mu=967.7$ MeV, $C_G\simeq-36.15~\mathrm{GeV}^{-2}$) is calibrated to the $5.2$ MeV pole and serves only as a regulator-sensitivity study. We do not reproduce the complex lattice pole or perform a full coupled-channel calculation.
\end{abstract}


\section{Introduction}
The spectroscopy of hadrons with more than three valence quarks has moved from a largely exploratory topic to a central part of contemporary hadron physics. The discoveries of the $X(3872)$, charged charmonium-like states, structures in the fully charmed sector, and the doubly charmed $T_{cc}^{+}$ have made clear that near-threshold peaks can emerge from more than one dynamical mechanism. Compact multiquark configurations, hadronic molecules, threshold cusps, and coupled-channel mixtures may all produce structures in nearby energy regions. The task is therefore not merely to assign a name to a peak, but to connect a microscopic description to observables: finite-volume spectra, scattering amplitudes, and poles on the appropriate Riemann sheets. General reviews of the exotic-hadron field emphasize precisely this plurality of possible structures and mechanisms \cite{ChenReview,GuoReview,BrambillaReview}.

The $T_{cc}^{+}$ is especially instructive because it lies extremely close to the $DD^*$ threshold. LHCb reported a narrow structure in the $D^0D^0\pi^+$ final state with valence content $cc\bar u\bar d$ and quantum numbers consistent with $I(J^P)=0(1^+)$ \cite{LHCbNature,LHCbNC}. For such a shallow state, small changes in the interaction or in the treatment of the $D^*$ instability can noticeably affect the inferred pole. A mass estimate alone does not determine whether the relevant pole is a bound state, a virtual state, or a resonance; the analytic structure of the amplitude is indispensable.

Two languages are commonly used to describe this system. Constituent and compact multiquark approaches organize the four constituents into colour, spin, flavour, and spatial configurations. In the diquark--antidiquark basis, the two charm quarks can couple to either $\bar{\mathbf 3}$ or $\mathbf 6$ in colour. Hadronic effective theories instead use colour-singlet mesons, such as $D$ and $D^*$, and describe their interaction through pion exchange, contact terms, and channel transitions. Detailed constituent-quark and coupled-channel studies can calculate spectra and transition dynamics, but their outputs depend on the chosen Hamiltonian, spatial ansatz, interaction parameters, and included channels \cite{Ortega2023,Qiu2024}. Lattice-QCD analyses provide a complementary route by relating finite-volume spectra to scattering amplitudes, while requiring careful treatment of the ensemble, channel content, left-hand singularities, and—in this case—the nearby $DD\pi$ channel \cite{Prelovsek2025,Whyte2025,Dawid2025,DawidCompare2025}. Neither language can simply be substituted for the other. A colour recoupling coefficient is not a scattering coupling, and a basis vector is not automatically a physical state or a pole.

The practical advantage of the colour-Casimir construction used here is narrower but well defined: it isolates, in a small and reproducible algebraic calculation, which pairwise colour coefficients contribute directly to the physical $DD^*$ channel and which combination controls transition to the orthogonal hidden-colour channel. This gives a transparent diagnostic that can be evaluated before choosing a spatial potential or fitting a hadronic interaction. It does not replace the dynamical information supplied by quark models, effective field theory, or lattice QCD; instead, it identifies the colour-algebra input and the additional dynamical kernel that a predictive calculation must supply.

The aim of this work is to make this boundary explicit while taking a concrete step toward a common framework. We derive the colour-Casimir matrices for a two-state singlet basis, write the colour recoupling to the physical $DD^*$ channel, and solve a minimal regulated Lippmann--Schwinger equation. The numerical example is deliberately labelled a calibration rather than a prediction. This makes the present result reproducible without hiding the central open problem: how to derive a momentum-dependent hadronic kernel from the graph-based colour structure.

The main results are as follows. First, the colour algebra yields an exact and compact criterion: the direct $DD^*$ element depends on two pair coefficients, and meson--hidden-colour mixing on a single anisotropy $\Delta_v$ (Secs.~\ref{sec:colour}--\ref{sec:hidden}). Second, this criterion is not sufficient for binding: in the Gaussian and constituent-model realizations studied here the hidden-colour level does not lie above the meson channel and the compact sector stays above threshold (Secs.~\ref{sec:spatial}--\ref{sec:matchedcomparison}). Third, a pion-mass test shows that colour algebra plus contact dynamics cannot reproduce the change from a bound to a virtual state between $m_\pi\simeq280$ and $391$ MeV, which locates the missing physics in pion exchange (Sec.~\ref{sec:pionmass}).

\section{Historical context and current status}
\subsection{Multiquark spectroscopy and the molecular question}
Multiquark configurations have been discussed since the early quark-model era. A classic question was whether heavy multiquark systems could form narrow or stable states rather than dissolving into two ordinary hadrons \cite{Ader}. Later work argued that heavy-quark dynamics could favour selected doubly heavy tetraquarks, particularly in the bottom sector \cite{KarlinerRosner,Francis2017,Leskovec2019}. The discovery of $X(3872)$ and other near-threshold states brought renewed attention to the distinction between compact and molecular descriptions. In a molecular picture, the state is dominated at long distances by two colour-singlet hadrons; in a compact picture, correlations among the constituent quarks are emphasized. In practice, these descriptions can mix, and their relative importance is often dependent on the scale and on the chosen interpolating basis.

For $T_{cc}^{+}$, the molecular interpretation is natural because of the proximity to $DD^*$ threshold. Yet proximity alone is not a proof of a purely molecular structure. The short-distance interaction, pion exchange, coupled channels, and the unstable nature of the $D^*$ all affect the amplitude. Effective field theory provides a systematic way to separate long-range contributions from contact interactions, while constituent models offer a different route to estimating the internal spectrum. Calculations including quark exchange and coupled hadronic channels have found that these effects can be significant in the charm sector \cite{Ortega2023,Qiu2024}. The different approaches should be compared through common observables, not by identifying their internal basis states one-to-one.

\subsection{Lattice QCD and the finite-volume spectrum}
Lüscher's quantization condition and related finite-volume methods provide a link between finite-volume spectra and scattering amplitudes \cite{Luscher}. A finite-volume level below a nominal threshold is not, by itself, enough to identify the pole type; one must reconstruct an amplitude and analytically continue it. This is particularly important when a nearby channel or a left-hand singularity affects the energy dependence.

Recent studies have sharpened the picture. Gil-Domínguez, Giachino, and Molina analysed the quark-mass dependence of the $T_{cc}$ pole and emphasized the importance of extrapolating lattice results to the physical pion mass \cite{Gil2025}. Prelovsek \emph{et al.} included both meson--meson and diquark--antidiquark interpolators in a lattice analysis of $DD^*$ scattering and treated the left-hand cut and partial-wave mixing \cite{Prelovsek2025}. For their ensemble with $m_\pi\simeq280$ MeV, the reported pole is
\begin{equation}
E_{\rm pole}-E_{\rm th}=-5.2^{+0.7}_{-0.8}-i\,6.3^{+2.4}_{-4.8}\ {\rm MeV}.
\label{eq:prelovsek}
\end{equation}
The inclusion of a diquark--antidiquark interpolator changes the variational basis used to determine the finite-volume spectrum; it does not, by itself, mean that an extra microscopic colour force has been inserted into the hadronic interaction kernel.

A separate lattice study by Whyte, Wilson, and Thomas extracted coupled $DD^*$--$D^*D^*$ amplitudes at $m_\pi\simeq391$ MeV and found a virtual-state pole below the $DD^*$ threshold as well as a second pole associated with the nearby closed channel \cite{Whyte2025}. These results complement the $m_\pi\simeq280$ MeV study but should not be merged into a single fit without accounting for the different ensembles and channel content. Three-body effects are also important: the relativistic $DD\pi$ formalism developed by Dawid, Romero-López, and Sharpe found that the subthreshold pole positions can depend strongly on the three-body interaction input \cite{Dawid2025}. Their subsequent comparison of integral-equation approaches clarifies the relation between two-body Lippmann--Schwinger descriptions and three-particle formulations \cite{DawidCompare2025}. Recent chiral calculations have also examined two-pion exchange in $DD^*$ interactions \cite{XuRen2025}. The dependence on the light-quark mass has been studied on the lattice at $m_\pi\simeq280$ MeV \cite{Padmanath2022} and in a quark-mass scan \cite{Collins2024}, near the physical point with the HAL~QCD method \cite{Lyu2023}, and in coupled-channel approaches that include the three-body $DD\pi$ cut \cite{Du2022}. These works are the natural benchmarks for the pion-mass test of Sec.~\ref{sec:pionmass}.

\subsection{Colour-Casimir graphs and the present study}
At the constituent level, the $cc\bar u\bar d$ system admits two independent colour-singlet couplings in a diquark--antidiquark basis. The pair operators $T_i\cdot T_j$ encode the SU(3) Casimir structure of each pair. A graph representation can organize which pairwise links contribute to diagonal matrix elements and which can mix colour configurations. However, an adjacency matrix is a representation of an assumed pairwise colour structure; it does not by itself determine spatial wavefunctions, spin interactions, or the momentum dependence of a hadronic scattering kernel.

This issue is directly connected to our earlier work on colour-Casimir graph methods for fully-heavy tetraquarks. In that study, the four constituents were represented by a weighted complete graph and the spectrum of a colour-weighted adjacency matrix was used to organize fully-heavy tetraquark levels \cite{MonemzadehTazimi2026}. That study also contains a brief diagnostic application to $T_{cc}^{+}$ in which the same colour-weighted adjacency matrix gives a lowest level near $3967.5$ MeV, about $93$ MeV above the measured mass, and the missing binding is attributed to hadronic dynamics outside a static colour graph. It does not derive the recoupling to the $DD^*$ colour singlet, the hidden-colour mixing criterion, the scattering benchmark, or the pion-mass test developed here. The present system differs in an important way: $T_{cc}^{+}$ lies at a heavy-meson threshold, so long-range hadronic dynamics and channel unitarity cannot be expected to emerge from a static colour graph alone. The earlier graph framework therefore motivates the colour bookkeeping, while the present paper asks what can be established when that bookkeeping is connected—carefully and explicitly—to a scattering equation.

Relative to a conventional constituent-quark calculation, the main advantage here is not a more precise mass spectrum: our colour projection is analytic, makes the assumed pairwise interaction coefficients visible, and yields a compact criterion for colour-channel mixing without first solving a four-body Schrödinger equation. Relative to a hadronic effective-field-theory or meson-exchange calculation, it offers a microscopic colour-space bookkeeping step, but it does not independently determine the contact terms, pion-exchange kernel, or low-energy constants. Relative to lattice QCD, it is inexpensive and analytically inspectable, but it cannot determine a pole or its uncertainty from first principles. These are complementary roles rather than a ranking of methods. In particular, the advantage of our formulation is the explicit separation of (i) exact colour algebra, (ii) model-dependent spatial projection, and (iii) hadronic scattering dynamics. The numerical LS example below demonstrates how the last step can be implemented reproducibly; because its coupling is fitted to a target binding energy, it is a benchmark of the equation and regulator dependence, not evidence that the graph alone predicts the observed pole.

\section{Scope and objectives}
The objectives are deliberately limited and testable:
\begin{enumerate}[leftmargin=1.8em]
\item derive the pairwise colour-Casimir matrices in a two-state colour-singlet basis;
\item express the physical $DD^*$ colour singlet as a recoupled combination of the diquark--antidiquark basis states;
\item solve a regulated separable $S$-wave Lippmann--Schwinger equation and quantify its regulator sensitivity;
\item test, without refitting, whether couplings fixed at one lattice ensemble reproduce the pole at another pion mass; and
\item distinguish analytical results and calibrated outputs from a genuine prediction of a graph-derived interaction.
\end{enumerate}
The last point is part of the method, not an afterthought. A model becomes predictive only when its graph-induced kernel is specified independently enough that its parameters can be constrained by data rather than by imposing the desired pole.

\section{Colour-singlet basis and pairwise Casimir matrices}
\label{sec:colour}
We consider $cc\bar u\bar d$ in the isoscalar $J^P=1^+$ channel. The two independent colour-singlet basis states are
\begin{align}
|A\rangle&=\left|(cc)_{\bar{\mathbf 3}}(\bar u\bar d)_{\mathbf 3}\right\rangle,\label{eq:A}\\
|S\rangle&=\left|(cc)_{\mathbf 6}(\bar u\bar d)_{\bar{\mathbf 6}}\right\rangle.\label{eq:S}
\end{align}
We use $T^a=\lambda^a/2$ for quarks and the conjugate generators for antiquarks. In the basis $(|A\rangle,|S\rangle)$, the like-particle pair matrices are
\begin{equation}
T_{12}=T_{34}=\begin{pmatrix}-\frac23&0\\[2pt]0&\frac13\end{pmatrix}.
\label{eq:like}
\end{equation}
For mixed quark--antiquark pairs, a consistent choice of relative phase gives
\begin{align}
T_{13}=T_{24}&=\begin{pmatrix}-\frac13&-\frac1{\sqrt2}\\[2pt]-\frac1{\sqrt2}&-\frac56\end{pmatrix},\label{eq:13}\\
T_{14}=T_{23}&=\begin{pmatrix}-\frac13&+\frac1{\sqrt2}\\[2pt]+\frac1{\sqrt2}&-\frac56\end{pmatrix}.\label{eq:14}
\end{align}
The signs of the off-diagonal entries depend on the phase convention for the basis states; physical quantities are unchanged if the basis and all matrix elements are transformed consistently.

Let the pairwise colour operator be
\begin{equation}
V_{\rm col}=\sum_{i<j}v_{ij}\,T_i\!\cdot T_j,
\label{eq:Vcol}
\end{equation}
where $v_{ij}$ represents the spatial or effective coefficient assigned to pair $ij$. The matrix elements are
\begin{align}
(V_{\rm col})_{AA}&=-\frac23(v_{12}+v_{34})-\frac13(v_{13}+v_{14}+v_{23}+v_{24}),\label{eq:VAA}\\
(V_{\rm col})_{SS}&=+\frac13(v_{12}+v_{34})-\frac56(v_{13}+v_{14}+v_{23}+v_{24}),\label{eq:VSS}\\
(V_{\rm col})_{AS}&=\frac1{\sqrt2}(-v_{13}+v_{14}+v_{23}-v_{24}).\label{eq:VAS}
\end{align}
Equation~\eqref{eq:VAS} is the principal analytical colour result. It shows that the mixing of the two colour configurations is controlled by differences among the mixed-pair coefficients. For example, if $v_{13}=v_{14}$ and $v_{23}=v_{24}$, the off-diagonal contribution vanishes. A nonzero mixing term therefore reflects a corresponding asymmetry in the pair coefficients or their spatial expectation values; it does not follow merely from the existence of two basis states.

As an algebraic check, if all six coefficients are equal, $v_{ij}=v$, then
\begin{equation}
V_{\rm col}=-\frac83v\,\mathbb{1}.
\label{eq:uniform}
\end{equation}
This follows from the total colour-singlet identity $\sum_{i<j}T_i\cdot T_j=\frac12[C_2(\mathbf1)-\sum_i C_2(i)]=-8/3$. The uniform limit checks the colour algebra but is not a realistic spatial potential for the tetraquark.

\section{Recoupling to the physical $DD^*$ channel}
A physical two-meson channel consists of two colour-singlet mesons. The distinction between meson--meson and hidden-colour components is established in the tetraquark literature; for example, Kim and Kim analysed the role of hidden-colour components in a tetraquark mixing model, although not the specific $T_{cc}^{+}$ scattering projection developed here \cite{Kim2024Hidden}. In the phase convention used above, the colour wavefunction of the $DD^*$ channel can be expressed in the diquark basis as
\begin{equation}
|DD^*;\mathbf1\otimes\mathbf1\rangle=\frac1{\sqrt3}|A\rangle+\sqrt{\frac23}|S\rangle.
\label{eq:recoupling}
\end{equation}
The orthogonal combination is a hidden-colour state in this recoupling scheme. Equation~\eqref{eq:recoupling} is a colour identity only. A complete physical state also requires spin and flavour recoupling, spatial wavefunctions, the appropriate exchange symmetry of the identical charm quarks, and coupling to the nearby hadronic channels.

\begin{figure}[t]
\centering
\begin{tikzpicture}[>=Latex,node distance=1.1cm,state/.style={draw,rounded corners,align=center,minimum width=3.6cm,minimum height=0.85cm}]
\node[state] (A) {$|A\rangle$: $\bar{\mathbf3}\otimes\mathbf3$};
\node[state,below=of A] (S) {$|S\rangle$: $\mathbf6\otimes\bar{\mathbf6}$};
\node[state,right=3.4cm of A,yshift=-0.55cm] (DD) {$|DD^*;\mathbf1\otimes\mathbf1\rangle$};
\draw[->,thick] (A.east)--node[above,sloped] {$1/\sqrt3$}(DD.west);
\draw[->,thick] (S.east)--node[below,sloped] {$\sqrt{2/3}$}(DD.west);
\end{tikzpicture}
\caption{Colour recoupling from the two diquark--antidiquark singlets to the physical meson--meson colour singlet. The arrows are basis amplitudes, not dynamical transition strengths.}
\label{fig:recoupling}
\end{figure}
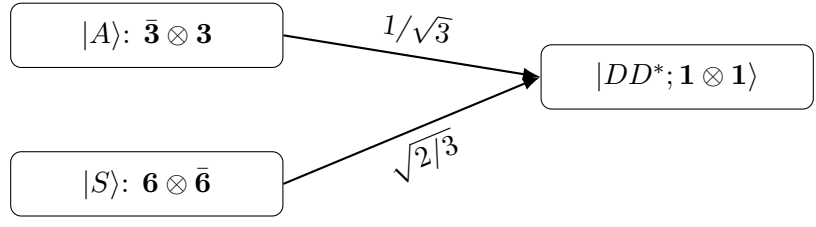

It would be too strong to read the coefficients in Eq.~\eqref{eq:recoupling} as probabilities for compact and molecular components. Basis coefficients depend on the representation and phase convention; a compositeness measure requires a dynamical definition and may be scheme dependent for a near-threshold state. Here, the recoupling is used only to show how the colour-singlet meson channel sits inside the four-quark colour space. Figure~\ref{fig:recoupling} summarizes the colour-basis amplitudes connecting the two diquark--antidiquark singlets to the physical $DD^*$ colour-singlet channel. These coefficients describe a change of basis, not a dynamical transition probability.

\section{A colour-neutrality result and the induced hidden-colour interaction}
\label{sec:hidden}
The recoupling in Eq.~\eqref{eq:recoupling} permits a direct test of what a pairwise colour-Casimir operator actually does in the physical meson channel. Define the normalized orthogonal hidden-colour combination
\begin{equation}
|H\rangle=\sqrt{\frac23}|A\rangle-\frac1{\sqrt3}|S\rangle,
\qquad \langle H|DD^*\rangle=0.
\label{eq:hidden}
\end{equation}
In the ordered basis $(|M\rangle,|H\rangle)$, where $|M\rangle\equiv|DD^*;\mathbf1\otimes\mathbf1\rangle$, the general pairwise operator of Eq.~\eqref{eq:Vcol} becomes
\begin{equation}
V_{MH}=\begin{pmatrix}V_{MM}&V_{MH}^{\rm off}\\ V_{MH}^{\rm off}&V_{HH}\end{pmatrix},
\label{eq:MHmatrix}
\end{equation}
with
\begin{align}
V_{MM}&=-\frac43(v_{13}+v_{24}),\label{eq:VMM}\\
V_{HH}&=-\frac13(v_{12}+v_{34})+\frac16(v_{13}+v_{24})-\frac76(v_{14}+v_{23}),\label{eq:VHH}\\
V_{MH}^{\rm off}&=\frac{\sqrt2}{3}\big[-v_{12}-v_{34}+v_{14}+v_{23}\big].\label{eq:VMH}
\end{align}
These expressions give a useful physical diagnostic. The direct expectation value in the colour-singlet meson--meson channel contains only the coefficients associated with the two internal meson pairs, $13$ and $24$. The cross-cluster pairs $12$, $14$, $23$, and $34$ do not contribute to $V_{MM}$ at first order. This is the colour-neutrality cancellation expected for two isolated colour-singlet mesons; it does not imply that all inter-hadron interactions vanish, because spatial overlap and virtual transitions to hidden-colour configurations can generate residual interactions.

The same projection identifies the precise source of pairwise hidden-colour mixing. It is controlled by the anisotropy combination
\begin{equation}
\Delta_v\equiv v_{14}+v_{23}-v_{12}-v_{34},\qquad
V_{MH}^{\rm off}=\frac{\sqrt2}{3}\Delta_v.
\label{eq:Deltav}
\end{equation}
Consequently, a uniform edge assignment $v_{ij}=v$ gives $V_{MH}^{\rm off}=0$ and $V_{MM}=V_{HH}=-8v/3$: the colour graph is proportional to the identity and cannot distinguish the two singlet configurations. More generally, if $\Delta_v=0$, this pairwise operator produces no direct mixing between the selected meson-singlet and hidden-colour states, even if other entries differ.

A further consequence follows when the hidden-colour configuration is higher in energy by $\Delta_H>0$ and its coupling is perturbative. Eliminating that state at second order gives the familiar level-repulsion correction
\begin{equation}
\delta E_M\simeq-\frac{|V_{MH}^{\rm off}|^2}{\Delta_H}
=-\frac{2}{9}\frac{(\Delta_v)^2}{\Delta_H}<0.
\label{eq:secondorder}
\end{equation}
Thus virtual hidden-colour mixing is attractive at this order only when the hidden-colour level lies above the meson channel, and its magnitude is quadratic in the spatial edge anisotropy. This is a sign and scaling result, not a numerical binding-energy prediction: $\Delta_v$ and $\Delta_H$ must be derived from a specified spatial Hamiltonian or constrained independently. In particular, the colour algebra alone does not fix them. Equation~\eqref{eq:secondorder} supplies a sharper criterion for any proposed graph-to-hadron model: it must generate a nonzero, independently justified $\Delta_v$ to obtain this mechanism of attraction.

\section{Spatial projection of the colour edges: a Gaussian diagnostic}
\label{sec:spatial}
The previous section isolated the colour algebra but left the edge coefficients $v_{ij}$ unspecified. To make their spatial origin explicit, we now evaluate a transparent Gaussian diagnostic. This is not yet a variational prediction for $T_{cc}^{+}$: its purpose is to determine which combinations of spatial scales enter the colour mixing and to test whether the assumption of a higher hidden-colour level is automatic.

Let particles $1,2$ be the charm quarks and $3,4$ the light antiquarks, and introduce Jacobi coordinates
\begin{equation}
\mathbf{x}=\mathbf{r}_1-\mathbf{r}_2,\quad
\mathbf{y}=\mathbf{r}_3-\mathbf{r}_4,\quad
\mathbf{z}=\frac{\mathbf{r}_1+\mathbf{r}_2}{2}-\frac{\mathbf{r}_3+\mathbf{r}_4}{2}.
\end{equation}
For a normalized, isotropic Gaussian probability density in each independent coordinate, take the variance of each Cartesian component to be $s_x^2$, $s_y^2$, and $s_z^2$, respectively. Then the pair separation $r_{ij}$ is distributed as a three-dimensional Gaussian with component variance $s_{ij}^2$. In particular,
\begin{align}
 s_{12}&=s_x, & s_{34}&=s_y,\\
 s_{13}=s_{14}=s_{23}=s_{24}&\equiv s_m
 =\sqrt{s_z^2+\frac{s_x^2+s_y^2}{4}}.
\end{align}
The equality of the four mixed-pair widths follows from this simple uncorrelated Gaussian ansatz; a more general correlated wavefunction need not preserve it. The needed expectation values are analytic:
\begin{equation}
 \left\langle\frac{1}{r_{ij}}\right\rangle=\sqrt{\frac{2}{\pi}}\frac{1}{s_{ij}},\qquad
 \langle r_{ij}\rangle=2\sqrt{\frac{2}{\pi}}s_{ij}.
 \label{eq:gaussianmoments}
\end{equation}
For a Coulombic pair coefficient $v_{ij}=\alpha_s\hbar c\langle 1/r_{ij}\rangle$, the edge anisotropy is therefore
\begin{equation}
 \Delta_v=\alpha_s\hbar c\sqrt{\frac{2}{\pi}}
 \left(\frac{2}{s_m}-\frac{1}{s_x}-\frac{1}{s_y}\right).
 \label{eq:spatialDeltav}
\end{equation}
Equation~\eqref{eq:spatialDeltav} is the requested spatial extraction: the mixing is controlled not by colour recoupling alone, but by the contrast between intra-cluster inverse radii and the inverse mixed-pair radius. For this Gaussian ansatz, the corresponding off-diagonal matrix element is $V_{MH}^{\rm off}=\sqrt{2}\Delta_v/3$.

As a scale illustration only, choose $s_x=0.30$ fm, $s_y=0.45$ fm, $s_z=1.00$ fm, and $\alpha_s=0.40$. These are illustrative widths, not parameters fitted to meson spectroscopy or lattice data. They give $s_m=1.036$ fm and, with $\hbar c=197.327$ MeV fm,
\begin{align}
 v_{12}&\simeq209.9~\mathrm{MeV}, & v_{34}&\simeq140.0~\mathrm{MeV},\\
 v_{13}=v_{14}=v_{23}=v_{24}&\simeq60.8~\mathrm{MeV}, &
 \Delta_v&\simeq-228.3~\mathrm{MeV},\\
 V_{MH}^{\rm off}&\simeq-107.6~\mathrm{MeV}.
\end{align}
The colour-potential diagonal entries from Eqs.~\eqref{eq:VMM}--\eqref{eq:VHH} are $V_{MM}\simeq-162.1$ MeV and $V_{HH}\simeq-238.2$ MeV, so the Coulomb-only diagonal difference is $V_{HH}-V_{MM}\simeq-76.1$ MeV. In this illustration, the hidden-colour diagonal lies lower, not higher. The hypothesis $\Delta_H>0$ used in Eq.~\eqref{eq:secondorder} is therefore not guaranteed by the colour algebra; it depends on the full spatial Hamiltonian, including kinetic energy, confinement, spin-dependent terms, and the appropriate channel boundary conditions. Moreover, $|V_{MH}^{\rm off}|/|V_{HH}-V_{MM}|\simeq1.41$, so second-order perturbation theory is not controlled for these illustrative values.

This diagnostic yields a useful negative result: \textbf{a nonzero colour-edge anisotropy does not by itself establish an attractive induced interaction or a bound $T_{cc}^{+}$ pole.} The sign of the second-order shift depends on the dynamically determined level ordering, and the magnitude depends on spatial wavefunctions. A defensible numerical prediction requires a common Hamiltonian, a variational solution for the spatial basis, and a coupled-channel calculation; the illustrative numbers above must not be quoted as a prediction for the physical state.

\section{First-stage variational check with a calibrated constituent Hamiltonian}
\label{sec:variational}
The preceding Gaussian edge estimate isolates the colour algebra but does not solve the spatial four-body problem. We therefore performed a separate first-stage variational diagnostic to test whether a restricted compact configuration is, by itself, driven below the model's two-meson threshold. This calculation uses an explicit constituent Hamiltonian and is kept distinct from both the illustrative Coulomb-only edge estimate and the separable hadronic benchmark discussed in Secs.~\ref{sec:ls} and~\ref{sec:benchmark}.

We use $T_i^a=\lambda_i^a/2$ for quarks and $T_i^a=-\lambda_i^{a*}/2$ for antiquarks, with the centre-of-mass kinetic energy removed:
\begin{equation}
 H_4=\sum_{i=1}^4m_i+T_{\rm rel}-\frac34\sum_{i<j}(\bm T_i\!\cdot\!\bm T_j)
 \left[V^C_{ij}(r_{ij})+V^{SS}_{ij}(r_{ij})\right].
 \label{eq:variationalH}
\end{equation}
Here $T_{\rm rel}$ denotes the nonrelativistic kinetic energy of the relative (Jacobi) motion with the centre-of-mass part removed; for a Gaussian of component variance $s^2$ it is $\langle T\rangle=3\hbar^2/(8\mu s^2)$, and the notation does not imply relativistic kinematics. The central and smeared spin--spin terms are
\begin{align}
 V^C_{ij}(r)&=-\frac{\kappa}{r}+\sigma r-C,\\
 V^{SS}_{ij}(r)&=\frac{A_{\rm hf}}{m_i m_j}
 \exp\!\left(-\frac{r^2}{r_{0,ij}^2}\right)\bm\sigma_i\!\cdot\!\bm\sigma_j,
 \label{eq:variationalV}
\end{align}
where $r_{0,ij}=[\alpha+\beta m_i m_j/(m_i+m_j)]^{-1}$. This is a phenomenological model variant specified here for a restricted diagnostic; its parameters are not claimed to be a faithful reproduction of a particular published Hamiltonian. We use $m_q=466$ MeV, $m_c=1933$ MeV, $\kappa=212$ MeV fm, $\sigma=940.946$ MeV/fm, $\alpha=1.09$ fm$^{-1}$, and $\beta=1.31\times10^{-4}$ (MeV fm)$^{-1}$. The constituent masses and potential parameters should therefore be read as model inputs, not as independently validated parameters for the physical $T_{cc}^{+}$ system. With $r_{0,cq}=0.877818$ fm, the hyperfine strength and additive constant are calibrated within this Gaussian implementation to the conventional $D$ and $D^*$ masses. The resulting values are $A_{\rm hf}=4.169235\times10^7$ MeV$^3$ and $C=734.5848$ MeV. The corresponding Gaussian widths are $s_D=0.267819$ fm and $s_{D^*}=0.277710$ fm, and the calibration gives $M_D=1869.65$ MeV and $M_{D^*}=2011.75$ MeV, so that
\begin{equation}
 E_{\rm th}^{\rm model}=M_D+M_{D^*}=3881.40~\mathrm{MeV}.
 \label{eq:variationalthreshold}
\end{equation}
These meson masses and the threshold are calibration outputs, not predictions.

For the compact sector, introduce the Jacobi coordinates
\begin{equation}
 \bm x=\bm r_1-\bm r_2,\quad \bm y=\bm r_3-\bm r_4,\quad
 \bm z=\frac{\bm r_1+\bm r_2}{2}-\frac{\bm r_3+\bm r_4}{2},
\end{equation}
and a product Gaussian proportional to $\exp[-x^2/(4s_x^2)-y^2/(4s_y^2)-z^2/(4s_z^2)]$. In the restricted Pauli-allowed $S$-wave colour--spin space, we retain
\begin{align}
 |A\rangle&=|(cc)_{\bar{\mathbf3},S=1}(\bar u\bar d)_{\mathbf3,S=0};J=1\rangle,\\
 |S\rangle&=|(cc)_{\mathbf6,S=0}(\bar u\bar d)_{\bar{\mathbf6},S=1};J=1\rangle.
\end{align}
Minimizing the lower eigenvalue of the $2\times2$ Rayleigh--Ritz matrix gives
\begin{equation}
 (s_x,s_y,s_z)=(0.216311,\,0.363405,\,0.250490)~\mathrm{fm},
\end{equation}
with the internal matrix (excluding constituent rest masses)
\begin{equation}
 H_{\rm comp}^{\rm int}=\begin{pmatrix}-756.968&67.887\\67.887&-351.444\end{pmatrix}~\mathrm{MeV}.
 \label{eq:compactmatrix}
\end{equation}
Adding $2m_c+2m_q=4798$ MeV gives the two restricted compact trial levels
\begin{equation}
 M_{\rm comp}^{(1)}=4029.97~\mathrm{MeV},\qquad
 M_{\rm comp}^{(2)}=4457.62~\mathrm{MeV}.
 \label{eq:compactlevels}
\end{equation}
The lowest compact trial level therefore lies $148.57$ MeV above the model threshold in Eq.~\eqref{eq:variationalthreshold}. This is a variational statement about the selected compact subspace only. It neither rules out a shallow molecular state nor establishes the absence of a near-threshold pole: the open $DD^*$ continuum and its correctly antisymmetrized transition kernels are not included in this compact-sector minimization.

In particular, the off-diagonal entry $67.887$ MeV in Eq.~\eqref{eq:compactmatrix} is an internal compact-basis matrix element, not the momentum-dependent transition form factor $V_{M\alpha}(p,p')$ required by a scattering equation. The latter must be evaluated between fully antisymmetrized meson--meson and compact rearrangement states, with their spatial overlap and spin--colour recoupling included. We therefore do not use this matrix element as an LS coupling and do not infer a pole from it. The calculation adds a useful negative diagnostic to the colour analysis: within this restricted ansatz, compact binding alone is insufficient to place the lowest trial level below threshold.

\subsection{A controlled common-width comparison with the colour--spin benchmark}
\label{sec:matchedcomparison}
We now perform a deliberately limited common-input test of whether the graph diagnostic supplies an additional prediction beyond the calibrated colour--spin calculation of Sec.~\ref{sec:variational}. The first requirement for a fair test is to avoid comparing matrix elements defined in different bases or generated by different operators as though they were the same observable. The compact colour--spin matrix in Eq.~\eqref{eq:compactmatrix} is expressed in the $(|A\rangle,|S\rangle)$ basis and contains central, confining, and spin--spin contributions. By contrast, $V_{MH}^{\rm off}$ is a colour-only Coulomb diagnostic in the recoupled $(|M\rangle,|H\rangle)$ basis. Their off-diagonal entries are therefore not directly comparable, and their signs can also depend on basis-phase conventions.

As a controlled geometry check, we evaluate the graph diagnostic at the same Gaussian widths that minimize the restricted colour--spin Rayleigh--Ritz matrix, $(s_x,s_y,s_z)=(0.216311,0.363405,0.250490)$ fm, keeping the graph-only coupling at the illustrative value $\alpha_s=0.40$. The mixed width is then $s_m=0.327809$ fm. The resulting Coulomb-only edge coefficients are $v_{12}=291.144$ MeV, $v_{34}=173.299$ MeV, and $v_m=192.117$ MeV, giving
\begin{equation}
 \Delta_v^{\rm matched}=2v_m-v_{12}-v_{34}=-80.209~\mathrm{MeV},\qquad
 V_{MH}^{\rm off,matched}=\frac{\sqrt2}{3}\Delta_v^{\rm matched}=-37.811~\mathrm{MeV}.
 \label{eq:matchedgraph}
\end{equation}
For comparison, the calibrated colour--spin calculation at its own variational minimum gives $H_{AS}^{\rm int}=67.887$ MeV in the $(|A\rangle,|S\rangle)$ basis [Eq.~\eqref{eq:compactmatrix}] and a lowest compact trial level $148.57$ MeV above its model $DD^*$ threshold. The shared-width evaluation is useful because it removes one source of arbitrariness in the spatial scales; it does not make the two off-diagonal elements equivalent. In particular, the graph value uses only the illustrative Coulomb edge prescription, whereas the reference matrix element contains the full model's central and spin--spin expectation values and is written in another basis.

To complete the operator-level comparison, we now extract the pair coefficients from the 
central part of the very same Hamiltonian in Eq.~\eqref{eq:variationalH}, using the optimized widths above. To match the convention in Eq.~\eqref{eq:Vcol}, define
\begin{equation}
 v_{ij}^{C}\equiv-\frac34\langle V^C_{ij}\rangle.
 \label{eq:vcdefinition}
\end{equation}
The resulting coefficients are $v_{12}^{C}=893.827$ MeV, $v_{34}^{C}=490.788$ MeV, and $v_{13}^{C}=v_{14}^{C}=v_{23}^{C}=v_{24}^{C}=568.783$ MeV. Thus
\begin{equation}
 \Delta_v^C=v_{14}^{C}+v_{23}^{C}-v_{12}^{C}-v_{34}^{C}
 =-247.050~\mathrm{MeV}.
 \label{eq:matchedcentralanisotropy}
\end{equation}
The central operator in the $(|A\rangle,|S\rangle)$ basis is diagonal for these equal mixed-pair coefficients. Including the $-3/4$ convention in Eq.~\eqref{eq:variationalH}, its matrix and the spin--spin contribution at the same widths are
\begin{equation}
 H_C^{AS}=\begin{pmatrix}-1681.454&0\\0&-1434.403\end{pmatrix}\!\mathrm{MeV},\qquad
 H_{SS}^{AS}=\begin{pmatrix}-182.821&67.887\\67.887&-24.347\end{pmatrix}\!\mathrm{MeV}.
 \label{eq:matchedASdecomp}
\end{equation}
The kinetic contribution in this restricted basis is proportional to the identity at the chosen widths. Equation~\eqref{eq:matchedASdecomp} therefore separates the origin of the compact off-diagonal element: it comes from the spin--spin operator in this calculation, not from a central colour-edge asymmetry between the mixed pairs.

For an algebraic common-basis check, use the orthogonal colour recoupling matrix
\begin{equation}
 U=\begin{pmatrix}1/\sqrt3&\sqrt{2/3}\\\sqrt{2/3}&-1/\sqrt3\end{pmatrix},\qquad
 H^{MH}=U^{T}H^{AS}U,
 \label{eq:matchedU}
\end{equation}
where the columns define $|M\rangle=(|A\rangle+\sqrt2|S\rangle)/\sqrt3$ and $|H\rangle=(\sqrt2|A\rangle-|S\rangle)/\sqrt3$, matching Eq.~\eqref{eq:hidden}. With this phase convention, the central and spin--spin matrices become
\begin{equation}
 H_C^{MH}=\begin{pmatrix}-1516.754&-116.461\\-116.461&-1599.104\end{pmatrix}\!\mathrm{MeV},\qquad
 H_{SS}^{MH}=\begin{pmatrix}-13.167&-52.077\\-52.077&-194.001\end{pmatrix}\!\mathrm{MeV}.
 \label{eq:matchedMHdecomp}
\end{equation}
The central off-diagonal element satisfies $H_{C,MH}=\sqrt2\Delta_v^C/3=-116.461$ MeV, as required by the colour projection. The spin--spin matrix is an additional operator structure and is not encoded by a scalar colour-only edge weight. Although Eq.~\eqref{eq:matchedU} is an exact two-state basis rotation, it is not by itself a full transformation to a physical $DD^*$ scattering channel: that interpretation also requires the spin and flavour recouplings, exchange symmetry, spatial overlaps, and continuum transition kernel.

This same-Hamiltonian check sharpens the conclusion. The central graph indicator reproduces a matrix element already contained in the central colour operator; it is a transparent re-expression of that operator, not an additional independent dynamical contribution. The spin--spin interaction carries information that a scalar colour graph alone does not capture. We cannot honestly compare an out-of-calibration physical quantity such as a pole position or finite-volume level using the present compact Hamiltonian, because it contains neither the $DD^*$ continuum nor the meson--compact transition kernel. The one-channel separable benchmark elsewhere in this paper is also not an independent test: its coupling was fitted to the target binding energy. We therefore make no claim of independent predictive power. The graph index is retained as an algebraically transparent diagnostic of colour anisotropy, while a physical prediction is left to a future, fully specified coupled-channel calculation.

\section{From a colour matrix to a hadronic kernel}
A key modelling step is to pass from a finite-dimensional colour matrix to a momentum-dependent hadronic interaction. The colour operator in Eq.~\eqref{eq:Vcol} acts in internal colour space, whereas a scattering kernel must depend on incoming and outgoing relative momenta and on the hadronic channel labels. A future graph-derived interaction might be written schematically as
\begin{equation}
V_{\alpha\beta}(p,p';E)=V_{\alpha\beta}^{(\pi)}(p,p';E)+V_{\alpha\beta}^{(\mathrm{short})}(p,p';\Lambda)+g_G F_{\alpha\beta}(p,p';\Lambda),
\label{eq:fullV}
\end{equation}
where $\alpha,\beta$ label hadronic channels, $V^{(\pi)}$ denotes pion-exchange contributions when relevant, $V^{(\mathrm{short})}$ represents short-range terms, and $F_{\alpha\beta}$ is the graph-inspired component. To make Eq.~\eqref{eq:fullV} predictive, one must specify the projection from colour to hadronic channels, the momentum dependence, and the dimensions and normalization of $g_GF$. One must also avoid double counting if static potentials inferred from Wilson loops or other effective interactions are already included.

A Wilson-loop correlation matrix and a colour adjacency matrix are not the same object. Wilson loops constrain static energies and can encode string breaking, while a graph matrix records an assumed pairwise colour structure. Combining them requires an explicit matching prescription. For this reason, we do not insert an arbitrary graph term and interpret its fitted coefficient as a prediction. Instead, the next section provides a transparent scattering benchmark whose limitations are identifiable.

\section{Separable $S$-wave Lippmann--Schwinger equation}\label{sec:ls}
For a reproducible baseline, take a single $DD^*$ channel and a Gaussian separable interaction
\begin{equation}
V(p,p')=C_G\,g(p)g(p'),\qquad g(p)=\exp\left(-\frac{p^2}{\Lambda^2}\right).
\label{eq:kernel}
\end{equation}
Here $C_G$ is an effective hadronic coupling and $\Lambda$ is the regulator scale. In this benchmark, $C_G$ is not derived from the colour matrices; it parametrizes the short-range interaction after projection into the physical channel. With the nonrelativistic free resolvent
\begin{equation}
G_0(E;q)=\frac{1}{E-q^2/(2\mu)+i0},
\end{equation}
the Lippmann--Schwinger equation reads
\begin{equation}
T(p,p';E)=V(p,p')+\int_0^\infty\frac{q^2dq}{2\pi^2}V(p,q)\frac{T(q,p';E)}{E-q^2/(2\mu)+i0}.
\label{eq:LS}
\end{equation}
Because the kernel is separable, set $T(p,p';E)=g(p)\tau(E)g(p')$. Substitution into Eq.~\eqref{eq:LS} gives the algebraic relation $\tau=C_G+C_G I(E)\tau$, where
\begin{equation}
I(E)=\frac{1}{2\pi^2}\int_0^\infty dq\,\frac{q^2e^{-2q^2/\Lambda^2}}{E-q^2/(2\mu)+i0}.
\label{eq:loop}
\end{equation}
Hence
\begin{equation}
T(p,p';E)=g(p)\,\frac{1}{C_G^{-1}-I(E)}\,g(p'),\qquad C_G^{-1}-I(E_{\rm pole})=0.
\label{eq:Tsol}
\end{equation}
This is the exact solution of the chosen separable equation up to evaluation of the one-dimensional loop integral. It is not a solution of a fully specified coupled-channel colour-graph model.

For a real energy below threshold and a real one-channel kernel, $I(E)$ is real and the bound-state pole lies on the real axis. A complex pole with nonzero imaginary part requires additional dynamical structure, such as open channels, left-hand-cut effects, or a consistent treatment of $DD\pi$. It cannot be generated simply by choosing a different real value of $C_G$ in this minimal kernel.

\section{Numerical benchmark and regulator dependence}\label{sec:benchmark}
We use $\mu=967.7$ MeV and $\Lambda=650$ MeV. To illustrate the solution, the coupling is fixed by imposing a binding energy $B=5.2$ MeV, or $E_{\rm target}=-5.2$ MeV relative to threshold. This target is the central real part of the lattice pole in Eq.~\eqref{eq:prelovsek}, not an independent output of the present model. Evaluation of Eq.~\eqref{eq:loop} gives
\begin{equation}
I(-5.2~\mathrm{MeV})\simeq -2.7662\times10^4~\mathrm{MeV}^2.
\label{eq:loopnum}
\end{equation}
The pole condition then fixes
\begin{equation}
C_G=\frac{1}{I(E_{\rm target})}\simeq -3.615\times10^{-5}~\mathrm{MeV}^{-2}=-36.15~\mathrm{GeV}^{-2}.
\label{eq:CGfit}
\end{equation}
Thus the chosen model has a pole at $E=-5.2$ MeV by construction. The reduced mass $\mu=967.7$ MeV corresponds to the physical $D$ and $D^*$ masses, whereas the target pole belongs to the $m_\pi\simeq280$ MeV ensemble, for which $\mu\simeq993$ MeV; this mismatch is removed in Sec.~\ref{sec:pionmass}, while the present benchmark is kept for continuity with Sec.~\ref{sec:variational}. The imaginary part of the published pole is not reproduced, and no independent fit to the finite-volume lattice spectrum is performed.

\subsection{Regulator scan}
\begin{figure}[t]
\centering
\begin{tikzpicture}
\begin{axis}[width=0.82\linewidth,height=6cm,xlabel={Regulator scale $\Lambda$ (MeV)},ylabel={Binding energy $B$ (MeV)},xmin=470,xmax=830,ymin=0,ymax=23,grid=major]
\addplot[mark=square*,thick] coordinates {(500,0.23)(600,2.65)(650,5.20)(700,8.90)(800,20.54)};
\end{axis}
\end{tikzpicture}
\caption{Binding-energy sensitivity when $C_G=-36.15\,\mathrm{GeV}^{-2}$ is held fixed. The variation illustrates why a fitting or renormalization prescription is needed.}
\label{fig:binding}
\end{figure}
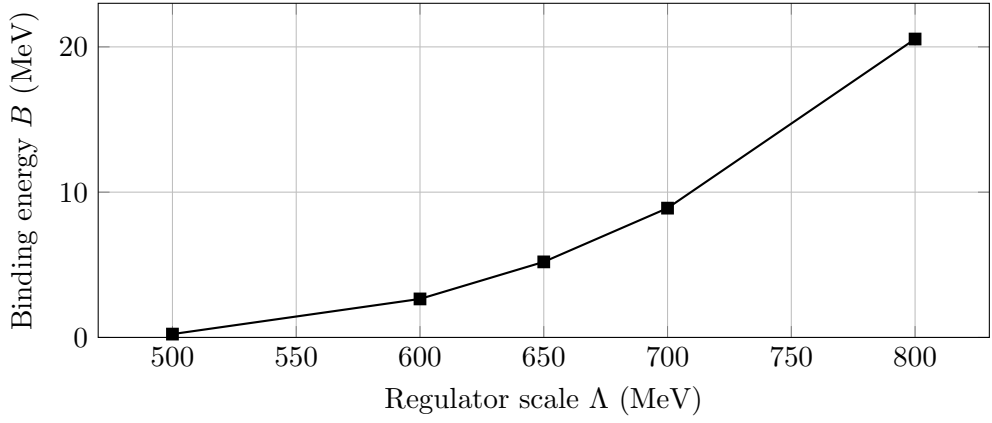

The regulator is not an observable. In a regulated effective interaction, the coupling generally changes with $\Lambda$ if a low-energy observable is held fixed. Table~\ref{tab:scan} summarizes the benchmark scan. The second column refits $C_G$ at each regulator to keep $B=5.2$ MeV, while the third holds $C_G=-36.15\,\mathrm{GeV}^{-2}$ fixed and shows the corresponding sensitivity of the binding energy. These approximate benchmark values are not statistical uncertainties and are not a fit to the full lattice spectrum. Figure~\ref{fig:coupling} shows the regulator dependence of the refitted coupling, while Fig.~\ref{fig:binding} shows the binding-energy sensitivity when that coupling is held fixed.

\begin{table}[t]
\centering
\caption{Illustrative regulator scan for the separable benchmark.}
\label{tab:scan}
\begin{tabular}{ccc}
\toprule
$\Lambda$ (MeV) & $C_G$ refitted ($\mathrm{GeV}^{-2}$) & $B$ at fixed $C_G$ (MeV)\\
\midrule
500 & $-52.09$ & $0.23$\\
600 & $-40.31$ & $2.65$\\
650 & $-36.15$ & $5.20$\\
700 & $-32.74$ & $8.90$\\
800 & $-27.50$ & $20.54$\\
\bottomrule
\end{tabular}
\end{table}

\begin{figure}[t]
\centering
\begin{tikzpicture}
\begin{axis}[width=0.82\linewidth,height=6cm,xlabel={Regulator scale $\Lambda$ (MeV)},ylabel={$C_G$ ($\mathrm{GeV}^{-2}$)},xmin=470,xmax=830,ymin=-57,ymax=-23,grid=major]
\addplot[mark=*,thick] coordinates {(500,-52.09)(600,-40.31)(650,-36.15)(700,-32.74)(800,-27.50)};
\end{axis}
\end{tikzpicture}
\caption{Coupling required to retain $B=5.2$ MeV as the regulator changes. The coupling is scheme dependent and should not be interpreted as an observable.}
\label{fig:coupling}
\end{figure}
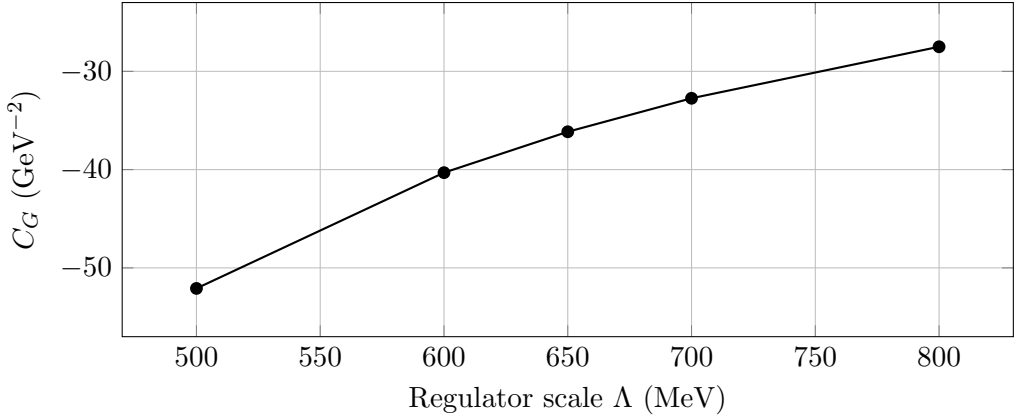

The scan has a straightforward interpretation. If the binding energy is fixed, changing the regulator can be compensated by refitting the short-range coupling. The coupling therefore changes while the calibrated energy remains fixed. If the coupling is held fixed, the binding energy changes substantially. This is expected for a regulated low-energy model and demonstrates why the number $C_G=-36.15\,\mathrm{GeV}^{-2}$ is not, by itself, a physical prediction.

\subsection{Sensitivity checks and the limits of calibration robustness}
\label{sec:sensitivity}
We now test whether the qualitative colour diagnostic depends on the illustrative Gaussian scales. Starting from $s_x=0.30$ fm, $s_y=0.45$ fm, $s_z=1.00$ fm, and $\alpha_s=0.40$, we vary each width independently by $\pm10\%$ and vary $\alpha_s$ by $\pm20\%$. This is a local stress test of the stated ansatz, not a statistical uncertainty estimate and not a scan over all possible four-body wavefunctions. Across the resulting corner points, the edge anisotropy remains negative, $-332.3\leq\Delta_v\leq-148.2$ MeV, and the mixing matrix element remains nonzero, $-156.7\leq V_{MH}^{\rm off}\leq-69.8$ MeV. Thus the existence and sign of the mixing are stable within this selected range. However, the diagonal gap also remains negative, $-110.8\leq V_{HH}-V_{MM}\leq-49.4$ MeV: the hidden-colour diagonal is still below the meson diagonal throughout this scan. Consequently, the illustrative parameter region does not satisfy the $\Delta_H>0$ condition required for the attractive second-order formula in Eq.~\eqref{eq:secondorder}. The robust conclusion is the presence of colour mixing in this ansatz, not a robust prediction of attraction or binding.

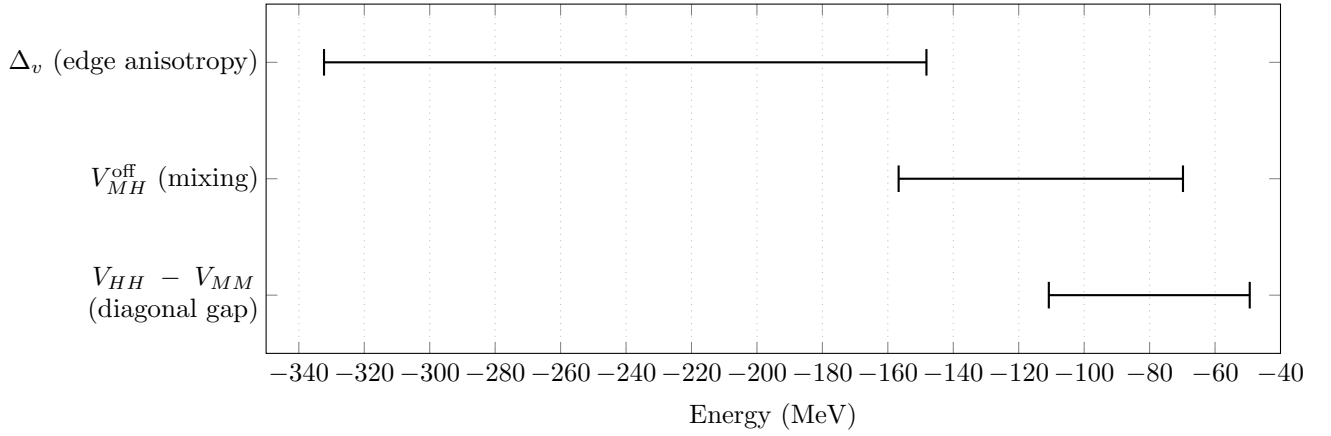
\begin{figure}[t]
\centering
\begin{tikzpicture}
\begin{axis}[
 width=0.92\linewidth,height=6.2cm,
 xmin=-350,xmax=-40,
 ymin=0.5,ymax=3.5,
 xlabel={Energy (MeV)},
 ytick={1,2,3},
 yticklabels={$\Delta_v$ (edge anisotropy), $V_{MH}^{\rm off}$ (mixing), $V_{HH}-V_{MM}$ (diagonal gap)},
 yticklabel style={align=right,text width=4.0cm},
 y dir=reverse,
 xmajorgrids=true,grid style={dotted},
 tick label style={font=\small},label style={font=\small},
 clip=false
]
\addplot[thick,mark=|,mark size=5pt] coordinates {(-332.3,1)(-148.2,1)};
\addplot[thick,mark=|,mark size=5pt] coordinates {(-156.7,2)(-69.8,2)};
\addplot[thick,mark=|,mark size=5pt] coordinates {(-110.8,3)(-49.4,3)};
\end{axis}
\end{tikzpicture}
\caption{Sensitivity envelopes from the independent corner scan in which each Gaussian width is varied by $\pm10\%$ and $\alpha_s$ by $\pm20\%$. Horizontal segments show the minimum-to-maximum ranges, not statistical confidence intervals. All ranges remain negative for this chosen ansatz; this supports stability of the sign within the scan, not a prediction of physical binding.}
\label{fig:colour-sensitivity}
\end{figure}

Figure~\ref{fig:colour-sensitivity} summarizes these corner-scan envelopes visually. The intervals remain entirely on the negative side for all three quantities, so the mixing sign and the lower hidden-colour diagonal are stable within this particular parameter envelope; the figure does not imply that the resulting interaction binds the physical $T_{cc}^{+}$.

\begin{table}[t]
\centering
\caption{Corner scan of the Gaussian colour-edge diagnostic. The widths are varied independently by $\pm10\%$ and $\alpha_s$ by $\pm20\%$ around the illustrative central values. These ranges are sensitivity envelopes, not confidence intervals.}
\label{tab:edgesensitivity}
\begin{tabular}{lrr}
\toprule
Quantity & Minimum (MeV) & Maximum (MeV)\\
\midrule
$\Delta_v$ & $-332.3$ & $-148.2$\\
$V_{MH}^{\rm off}$ & $-156.7$ & $-69.8$\\
$V_{HH}-V_{MM}$ & $-110.8$ & $-49.4$\\
\bottomrule
\end{tabular}
\end{table}

\subsection{Numerical reproducibility audit: direct matrix assembly and dense scan}
\label{sec:reproducibility}
To test the colour projection independently of the closed-form expressions used in the main derivation, we implemented a separate numerical check that constructs the full $2\times2$ operator from the six pairwise matrices $T_i\!\cdot T_j$, rotates it into the $(M,H)$ meson/hidden-colour basis, and compares its entries with the analytic expressions for $V_{MM}$, $V_{MH}^{\rm off}$, and $V_{HH}$ in Eqs.~(\ref{eq:VMM}), (\ref{eq:VMH}), and (\ref{eq:VHH}). The test used 10,000 pseudorandom sets of six edge coefficients drawn uniformly from $[-500,500]$ MeV with a fixed seed, so the calculation is exactly repeatable. The largest absolute discrepancies were $6.82\times10^{-13}$ MeV for $V_{MM}$, $3.98\times10^{-13}$ MeV for $V_{MH}^{\rm off}$, and $4.55\times10^{-13}$ MeV for $V_{HH}$. The uniform-edge identity, $\sum_{i<j}T_i\!\cdot T_j=-\frac{8}{3}\mathbf{1}$ in the chosen colour-singlet subspace, was satisfied to a maximum matrix-element error of $4.44\times10^{-16}$. These residuals are consistent with floating-point roundoff and provide a numerical check of the matrix recoupling and the stated analytic formulae.

We also replaced the finite corner scan by a deterministic dense grid with 101 points for each of $s_x$, $s_y$, and $s_z$, each width independently spanning $\pm10\%$ around its central value. This gives $101^3=1,030,301$ width triples. Since all three Coulomb-only diagnostics scale linearly with $\alpha_s$, evaluating the two endpoints $\alpha_s=0.32$ and $0.48$ captures the extrema in that variable as well. The resulting envelopes are shown in Table~\ref{tab:densegrid}. They reproduce the earlier corner-scan ranges to the displayed precision and remain entirely negative. In particular, the mixing element remains nonzero throughout this prescribed domain, while $V_{HH}-V_{MM}<0$ throughout it; the scan therefore does not enter the regime in which the positive-gap, second-order attractive-shift formula applies.

\begin{table}[t]
\centering
\caption{Results of the numerical reproducibility audit. The first three rows report maximum absolute differences between direct matrix assembly and the closed-form projections over 10,000 seeded random edge assignments. The final rows give the extrema over the dense Gaussian-width grid and both endpoints of $\alpha_s$. These are numerical verification and model-sensitivity results, not statistical uncertainties or physical predictions.}
\label{tab:densegrid}
\begin{tabular}{lrr}
\toprule
Quantity & Lower bound / error (MeV) & Upper bound / max. error (MeV)\\
\midrule
Maximum absolute error in $V_{MM}$ & --- & $6.82\times10^{-13}$\\
Maximum absolute error in $V_{MH}^{\rm off}$ & --- & $3.98\times10^{-13}$\\
Maximum absolute error in $V_{HH}$ & --- & $4.55\times10^{-13}$\\
$\Delta_v$ (MeV) & $-332.340$ & $-148.151$\\
$V_{MH}^{\rm off}$ (MeV) & $-156.667$ & $-69.839$\\
$V_{HH}-V_{MM}$ (MeV) & $-110.780$ & $-49.384$\\
\bottomrule
\end{tabular}
\end{table}

The audit establishes numerical consistency of the colour algebra and robustness of the signs within the selected Gaussian parameter box. It does not independently validate the Gaussian ansatz, the Coulomb-only interaction, the constituent parameters, or the mapping from this compact colour basis to the physical scattering amplitude. In particular, the graph diagnostic is a transparent combination of the same pairwise coefficients rather than an additional independent dynamical input. The reproducibility script and machine-readable scan output are provided as supplementary computational files, together with the scripts for the constituent-model calibration, the explicit colour--spin matrices, and the pion-mass scan of Sec.~\ref{sec:pionmass}; the main conclusions do not rely on interpreting this numerical check as evidence for binding.

A separate calibration check varies the imposed binding target over the interval suggested by the quoted real-part uncertainty, $B=4.4$--$5.9$ MeV, while keeping $\Lambda=650$ MeV and $\mu=967.7$ MeV. The corresponding fitted couplings are approximately $C_G=-35.15$ GeV$^{-2}$ at $B=4.4$ MeV, $-36.15$ GeV$^{-2}$ at $B=5.2$ MeV, and $-36.98$ GeV$^{-2}$ at $B=5.9$ MeV. This confirms that the fitted coupling is sensitive to the calibration target. It does not demonstrate that the pole prediction is robust: the pole is placed at each chosen target by construction. The existing fixed-coupling regulator scan in Table~\ref{tab:scan} is the more informative warning: changing the regulator while holding $C_G$ fixed changes the binding energy substantially. A genuine robustness test would require fitting a common set of finite-volume levels or amplitudes, with their covariance, and then predicting an independent observable or ensemble.

The colour--spin comparison is also deliberately limited. In Sec.~\ref{sec:variational}, the calibrated constituent Hamiltonian includes central and spin--spin terms and gives a lowest level $148.57$ MeV above its own model $DD^*$ threshold in the restricted compact Gaussian basis. This is a useful counterpoint to the colour-only diagnostic: adding spin-dependent dynamics does not, in this particular restricted variational calculation, produce compact binding. But because the continuum and meson--compact transition kernel are not included, it cannot be read as a prediction that the physical $T_{cc}^{+}$ is unbound. The $67.887$ MeV compact-basis off-diagonal element is not the scattering transition form factor. A fair quantitative comparison with a full colour--spin or coupled-channel model would require matching the basis, Hamiltonian, threshold, and fitted observables, which is beyond the present benchmark.

\section{Pion-mass dependence as a non-circular test}
\label{sec:pionmass}
The calibration of Sec.~\ref{sec:benchmark} places the pole at the chosen energy by construction. A test that is not circular must fix the couplings at one ensemble and then change only those inputs that the constituent model can actually predict. Two lattice analyses provide the required contrast. At $m_\pi\simeq280$ MeV, Prelovsek \emph{et al.} report a bound-state pole at $-5.2$ MeV \cite{Prelovsek2025}. At $m_\pi\simeq391$ MeV, Whyte, Wilson, and Thomas find instead a virtual state about $62$ MeV below the $DD^*$ threshold \cite{Whyte2025}. The two results are used here only as separate numbers; no joint fit is attempted, and the ensembles differ in lattice action, charm-quark tuning, and channel content. The test therefore addresses a qualitative question---whether the model can turn a shallow bound state into a deep virtual state when the light-quark mass changes---and not a precise pole extraction.

\subsection{Hidden-colour state as a discrete channel}
To keep the extension minimal, a discrete state $|H\rangle$ at energy $\Delta_H$ above the $DD^*$ threshold is added to the single-channel kernel of Eq.~\eqref{eq:kernel}, with a separable coupling $\lambda\,g(p)$ to the meson channel. Eliminating $|H\rangle$ gives an energy-dependent effective interaction,
\begin{equation}
V_{\rm eff}(E)=C_G+\frac{\lambda^2}{E-\Delta_H},\qquad
\frac{1}{V_{\rm eff}(E_{\rm pole})}-I(E_{\rm pole})=0 .
\label{eq:Veff}
\end{equation}
For $E<\Delta_H$ the second term is attractive; this is the non-perturbative counterpart of Eq.~\eqref{eq:secondorder}. The probability of the discrete component in a bound state is
\begin{equation}
Z_H=\frac{\lambda^2/(E-\Delta_H)^2}{\lambda^2/(E-\Delta_H)^2+J(E)/I(E)^2},\qquad
J(E)=\frac{1}{2\pi^2}\int_0^\infty dq\,\frac{q^2e^{-2q^2/\Lambda^2}}{\big(E-q^2/2\mu\big)^2}.
\label{eq:ZH}
\end{equation}
Writing $E=-\kappa^2/(2\mu)$, the loop integral has the closed form
\begin{equation}
I(\kappa)=-\frac{\mu}{\pi^2}\left[\frac{\Lambda}{2}\sqrt{\frac{\pi}{2}}-\frac{\pi\kappa}{2}\,e^{2\kappa^2/\Lambda^2}\,\mathrm{erfc}\!\left(\frac{\sqrt2\,\kappa}{\Lambda}\right)\right],
\label{eq:Ikappa}
\end{equation}
which is analytic in $\kappa$. A bound state corresponds to a root with $\kappa>0$ and a virtual state to a root with $\kappa<0$ (second sheet). For $\mu=967.7$ MeV, $\Lambda=650$ MeV and $E=-5.2$ MeV this reproduces Eq.~\eqref{eq:loopnum}.

The two parameters $C_G$ and $\lambda^2$ are not derived from the colour algebra: the overlap factor that would connect $V_{MH}^{\rm off}$ to $\lambda$ is not known. They are instead parametrized by $f=C_G/C_G^{(1)}$ with $C_G^{(1)}=1/I(E_{\rm pole})$ the one-channel value, so that $\lambda^2=\big(1/I-C_G\big)(E_{\rm pole}-\Delta_H)$ is fixed by the pole condition at the first ensemble. The prediction at the second ensemble is then a function of $f$ alone.

\subsection{Ensemble-consistent inputs}
The only quantities that change between ensembles are the reduced mass $\mu$ and the gap $\Delta_H$. Both are obtained from the constituent Hamiltonian of Sec.~\ref{sec:variational} with all potential parameters ($A_{\rm hf}$, $C$, $\kappa$, $\sigma$, $\alpha$, $\beta$) kept fixed and the two constituent masses $(m_q,m_c)$ refitted to the lattice $D$ and $D^*$ masses of each ensemble. The compact-sector gap is then recomputed with the same $2\times2$ Rayleigh--Ritz procedure. The $m_\pi\simeq280$ MeV values $m_D=1927$ MeV and $m_{D^*}=2049$ MeV are taken from Ref.~\cite{Prelovsek2025}. For $m_\pi\simeq391$ MeV, the values $a_tm_D=0.33281$, $a_tm_{D^*}=0.35464$ and $a_t^{-1}=5667$ MeV of Ref.~\cite{Whyte2025} give $m_D\simeq1886$ MeV and $m_{D^*}\simeq2010$ MeV. The inputs and results are collected in Table~\ref{tab:ensembles}. The kinetic energy used throughout is the nonrelativistic expression $\langle T\rangle=\tfrac{3}{8}\hbar^2\sum_i(\mu_is_i^2)^{-1}$ in the Jacobi coordinates of Eq.~\eqref{eq:variationalH}; with this form the calibration of Sec.~\ref{sec:variational} is reproduced exactly ($A_{\rm hf}=4.169235\times10^7$ MeV$^3$, $C=734.585$ MeV, $s_D=0.267819$ fm), as are the matrices of Eqs.~\eqref{eq:compactmatrix} and~\eqref{eq:matchedASdecomp}.

\begin{table}[t]
\centering
\caption{Ensemble-dependent inputs of the constituent model. $(m_q,m_c)$ are refitted to $(m_D,m_{D^*})$ with all other parameters fixed; only $\mu$ and $\Delta_H$ enter the scattering equation. The individual values of $m_q$ and $m_c$ come from a two-parameter fit and carry no independent physical meaning.}
\label{tab:ensembles}
\begin{tabular}{lrrrrrr}
\toprule
Ensemble & $m_D$ & $m_{D^*}$ & $\mu$ & $m_q$ & $m_c$ & $\Delta_H$\\
 & (MeV) & (MeV) & (MeV) & (MeV) & (MeV) & (MeV)\\
\midrule
Physical-mass model (Sec.~\ref{sec:variational}) & 1869.65 & 2011.75 & 969.05 & 466.00 & 1933.00 & 148.57\\
$m_\pi\simeq280$ MeV & 1927 & 2049 & 993.06 & 552.66 & 1960.14 & 213.87\\
$m_\pi\simeq391$ MeV & 1886.0 & 2009.7 & 972.94 & 557.80 & 1916.42 & 222.05\\
\bottomrule
\end{tabular}
\end{table}

\subsection{Result: the pole barely moves}
For $m_\pi\simeq280$ MeV with $\mu=993.06$ MeV one finds $I(-5.2~\mathrm{MeV})=-2.826\times10^4~\mathrm{MeV}^2$ and $C_G^{(1)}=-35.39~\mathrm{GeV}^{-2}$. This differs from Eq.~\eqref{eq:CGfit} because Sec.~\ref{sec:benchmark} uses the physical-mass reduced mass $\mu=967.7$ MeV; the benchmark of Sec.~\ref{sec:benchmark} is kept unchanged for continuity with Sec.~\ref{sec:variational}. Fixing $(C_G,\lambda^2)$ at the $280$ MeV pole and changing only $(\mu,\Delta_H)$ to their $391$ MeV values gives the results of Table~\ref{tab:pionmass}.

\begin{table}[t]
\centering
\caption{Pole at the $m_\pi\simeq391$ MeV ensemble predicted from couplings fixed at the $m_\pi\simeq280$ MeV pole ($E=-5.2$ MeV), for $\Lambda=650$ MeV. $f=C_G/C_G^{(1)}$ is the fraction of the attraction carried by the direct contact term; $\kappa=\sqrt{-2\mu E}$. The lattice result of Ref.~\cite{Whyte2025} is a virtual state at $E\simeq-62$ MeV ($\kappa\simeq-347$ MeV).}
\label{tab:pionmass}
\begin{tabular}{ccccc}
\toprule
$f$ & $C_G$ ($\mathrm{GeV}^{-2}$) & $\lambda^2$ ($\mathrm{MeV}^{-1}$) & $\kappa$ (MeV) & $E$ (MeV)\\
\midrule
0.00 & $0$ & $7.75\times10^{-3}$ & $+87$ & $-3.9$ (bound)\\
0.25 & $-8.85$ & $5.81\times10^{-3}$ & $+89$ & $-4.1$ (bound)\\
0.50 & $-17.69$ & $3.88\times10^{-3}$ & $+91$ & $-4.3$ (bound)\\
0.75 & $-26.54$ & $1.94\times10^{-3}$ & $+93$ & $-4.5$ (bound)\\
0.90 & $-31.85$ & $7.75\times10^{-4}$ & $+95$ & $-4.6$ (bound)\\
\midrule
\multicolumn{3}{l}{Lattice, $m_\pi\simeq391$ MeV \cite{Whyte2025}} & $\simeq-347$ & $\simeq-62$ (virtual)\\
\bottomrule
\end{tabular}
\end{table}

For every value of $f$ the model keeps a shallow bound state at about $-4$ to $-4.6$ MeV, only slightly less bound than at $280$ MeV. The result is insensitive to the regulator: for $f=0.5$ it is $-4.5$, $-4.3$, and $-4.1$ MeV for $\Lambda=500$, $650$, and $800$ MeV. The two ensemble-dependent inputs contribute comparably: for $f=0.5$, changing only $\mu$ gives $-4.7$ MeV, changing only $\Delta_H$ gives $-4.7$ MeV, and changing both gives $-4.3$ MeV. To reach the lattice virtual state, the effective interaction at $E=-62$ MeV would have to be about $0.16$ of the value the model produces there, that is, roughly $84\%$ weaker.

This is a negative but informative result. The heavy-quark scales that enter through $\mu$ and $\Delta_H$ change by only a few per cent between the two ensembles, so a model built from colour algebra, a contact interaction, and a compact-sector gap cannot generate the strong light-quark-mass dependence seen on the lattice. In the effective-field-theory and coupled-channel literature this dependence is associated with long-range pion exchange, the $D^*\to D\pi$ cut, and the left-hand singularity \cite{Padmanath2022,Du2022,Collins2024,Gil2025,Dawid2025,XuRen2025}, none of which is present in the kernel used here. The conclusion is conditional: the contact couplings were taken to be independent of $m_\pi$, and $\Delta_H$ is the gap of the restricted compact sector rather than of a fully antisymmetrized hidden-colour channel. Allowing $C_G$ or $\lambda$ to vary with $m_q$ would add free parameters and turn the test back into a fit. What survives is a concrete statement: a graph- or colour-based kernel that aims to be predictive must contain an explicit light-quark-mass dependence, most naturally through the pion-exchange part $V^{(\pi)}$ of Eq.~\eqref{eq:fullV}.

\subsection{Hidden-colour admixture}
For the physical-mass benchmark ($\mu=967.7$ MeV, $\Delta_H=148.57$ MeV, pole at $-5.2$ MeV) Eq.~\eqref{eq:ZH} gives $Z_H\simeq16\%$ when the whole attraction is carried by the hidden-colour exchange ($f=0$), and smaller values as the direct term takes over: $12.7\%$, $8.8\%$, and $4.6\%$ for $f=0.25$, $0.5$, and $0.75$. These numbers are upper bounds within the model, not measurements of compactness. They depend on the identification of $\Delta_H$ with the lowest compact level and on the unknown overlap that fixes $\lambda$; for the same reason they should not be read as compositeness probabilities.

\section{Discussion: what the benchmark does and does not show}
The hidden-colour decomposition itself is not new; it is a standard feature of multiquark colour recoupling. The result derived here is a channel-specific algebraic projection of a general pairwise operator. Its novelty is limited to making the cancellation and mixing criterion explicit in this basis; it is not, by itself, a new dynamical prediction or a complete graph-to-scattering construction. First, the colour algebra is explicit in the chosen basis. Second, after recoupling to the physical meson basis, the direct $DD^*$ matrix element depends only on the two intra-meson coefficients, displaying the cancellation of direct cross-cluster colour terms for colour-singlet mesons. Third, the meson--hidden-colour transition is proportional to $\Delta_v=v_{14}+v_{23}-v_{12}-v_{34}$; uniform edges give a multiple of the identity and no mixing. When the hidden-colour level is higher, its perturbative elimination produces the attractive shift $-2(\Delta_v)^2/(9\Delta_H)$. Fourth, the selected separable Lippmann--Schwinger equation is solved and its regulator sensitivity is displayed.

The numerical pole energy is not a new prediction. It is obtained by choosing $C_G$ so that the pole occurs at the central real part of a published lattice result. In addition, the one-channel real kernel does not reproduce the imaginary part of that pole. These limitations matter because the recent lattice and EFT literature shows that left-hand cuts, partial-wave mixing, coupled channels, pion-mass dependence, and $DD\pi$ dynamics can affect the analytic continuation \cite{Prelovsek2025,Whyte2025,Gil2025,Dawid2025,DawidCompare2025,XuRen2025}.

The comparison with established methods clarifies the appropriate claim of novelty. A constituent-quark or compact-tetraquark model can incorporate spin-dependent forces and spatial correlations and can therefore calculate a spectrum once its Hamiltonian is specified; a coupled-channel meson model can address thresholds and channel transitions directly \cite{Ortega2023,Qiu2024}. Lattice calculations can constrain scattering amplitudes and pole locations from finite-volume information, although the extraction depends on the simulated pion mass, operator basis, and analytic treatment of nearby singularities \cite{Prelovsek2025,Whyte2025}. The present colour-Casimir analysis is cheaper and more transparent at the algebraic stage: it shows that the direct colour-singlet matrix element is fixed by the two intra-meson pair coefficients, whereas the off-diagonal hidden-colour coupling is controlled by the single anisotropy combination $\Delta_v$. This compact criterion is the method's principal advantage. It can guide or constrain the construction of a more complete kernel, and it makes the origin of the mixing term easy to audit. It cannot, without additional spatial, spin, flavour, and momentum-dependent input, outperform those approaches in predicting a physical mass, width, or complex pole.

The independent compact-sector diagnostic in Sec.~\ref{sec:variational} provides a further constraint on interpretation: for the stated calibrated Hamiltonian and restricted Gaussian basis, the lowest compact trial level is about $149$ MeV above the model $DD^*$ threshold. This is not evidence against a molecular pole, because the open channel and its transition kernels were not solved. The compact off-diagonal matrix element cannot be substituted for the missing meson--compact kernel. The pion-mass test of Sec.~\ref{sec:pionmass} adds a quantitative version of this statement: with couplings fixed at $m_\pi\simeq280$ MeV, the model predicts a shallow bound state at $m_\pi\simeq391$ MeV where the lattice finds a virtual state about $62$ MeV below threshold, so a predictive kernel needs explicit light-quark-mass dependence. A predictive graph-based extension would need to define $F_{\alpha\beta}(p,p')$ before fitting, justify its projection from colour space to hadronic channels, include spin and flavour recoupling, and fit a single well-defined finite-volume data set with uncertainties and correlations. Data from lattice calculations at different pion masses should not be combined as if they came from the same ensemble; Sec.~\ref{sec:pionmass} therefore uses the two ensembles only as separate inputs, without a joint fit, and tests a qualitative trend (bound versus virtual) rather than extracting a pole. If Wilson-loop information is used to construct static interactions, the matching to the hadronic kernel must also avoid double counting interactions already represented by contact terms.

\section{Conclusions}
We have established three results for the $T_{cc}^{+}$ channel and delimited what they imply.

\emph{First}, colour algebra. In the physical $DD^*$ colour basis the direct pairwise matrix element is $V_{MM}=-4(v_{13}+v_{24})/3$ and the transition to the orthogonal hidden-colour configuration is $V_{MH}^{\rm off}=\sqrt2\,\Delta_v/3$, with $\Delta_v=v_{14}+v_{23}-v_{12}-v_{34}$. A uniform graph is proportional to the identity and cannot split or mix the two configurations. The projection is exact in the chosen two-state basis and agrees with direct matrix assembly to floating-point precision. The same-Hamiltonian decomposition of Sec.~\ref{sec:matchedcomparison} shows that the graph index reproduces the central operator's recoupled off-diagonal element ($116.46$ MeV), whereas the $67.887$ MeV compact-basis element originates in the spin--spin term. The graph index is therefore a transparent re-expression of the colour operator, not an independent dynamical prediction.

\emph{Second}, two negative dynamical results. In the Coulomb-type Gaussian diagnostic the hidden-colour diagonal lies below the meson diagonal throughout the sensitivity box ($-110.8\le V_{HH}-V_{MM}\le-49.4$ MeV), so the attractive second-order shift $-2\Delta_v^2/(9\Delta_H)$ is not realised there. In the calibrated constituent Hamiltonian the lowest compact trial level lies $148.57$ MeV above its model $DD^*$ threshold. This concerns the restricted compact sector and does not exclude a molecular pole.

\emph{Third}, the pion-mass test of Sec.~\ref{sec:pionmass}, which is not circular. With $C_G$ and the hidden-colour coupling fixed at the $m_\pi\simeq280$ MeV pole and only $\mu$ and $\Delta_H$ changed to their $m_\pi\simeq391$ MeV values, the model gives a bound state near $-4$ to $-4.6$ MeV for any split between direct and hidden-colour attraction and any regulator in $500$--$800$ MeV, while the lattice finds a virtual state near $-62$ MeV. The interaction would have to be about $84\%$ weaker. Light-quark-mass dependence must therefore enter the kernel explicitly, most naturally through pion exchange, the $D^*\to D\pi$ cut, and the left-hand singularity. This is the main constraint on any graph-derived hadronic kernel: it must contain $V^{(\pi)}$ of Eq.~\eqref{eq:fullV} before it can be predictive.

\emph{Scope.} The one-channel separable benchmark with $C_G\simeq-36.15~\mathrm{GeV}^{-2}$ places a pole at $-5.2$ MeV by construction, does not reproduce the imaginary part of the lattice pole, and serves only to document regulator sensitivity. The pion-mass conclusion assumes $m_\pi$-independent contact couplings and uses the compact-sector gap as $\Delta_H$; the two lattice ensembles are used as separate inputs, without a joint fit. A full coupled-channel calculation with antisymmetrised meson--compact kernels, spin and flavour recoupling, and $DD\pi$ dynamics remains future work. Relative to constituent-quark models and effective theories, the present method is cheaper and algebraically transparent but omits much of their spin and spatial dynamics; relative to lattice QCD it cannot determine a pole from first principles. Its value is methodological: the exact colour coefficients, the anisotropy driving channel mixing, the calibration, and the missing dynamical ingredients are kept separately identifiable.

\section*{Acknowledgements}
The authors acknowledge the University of Kashan for supporting the research environment in which this work was developed.

\section*{Conflict of interest}
The authors declare no conflict of interest.


\begin{thebibliography}{99}
\bibitem{Ader} J.-P. Ader, J.-M. Richard, and P. Taxil, ``Do narrow heavy multiquark states exist?'' \emph{Phys. Rev. D} \textbf{25}, 2370 (1982).
\bibitem{KarlinerRosner} M. Karliner and J. L. Rosner, ``Discovery of doubly-charmed $\Xi_{cc}$ baryon implies a stable $bb\bar u\bar d$ tetraquark,'' \emph{Phys. Rev. Lett.} \textbf{119}, 202001 (2017), \href{https://arxiv.org/abs/1707.07666}{arXiv:1707.07666}.
\bibitem{Francis2017} A. Francis, R. J. Hudspith, R. Lewis, and K. Maltman, ``Lattice prediction for deeply bound doubly heavy tetraquarks,'' \emph{Phys. Rev. Lett.} \textbf{118}, 142001 (2017), \href{https://arxiv.org/abs/1607.05214}{arXiv:1607.05214}.
\bibitem{Leskovec2019} L. Leskovec, S. Meinel, M. Pflaumer, and M. Wagner, ``Lattice QCD investigation of a doubly-bottom $\bar b\bar b ud$ tetraquark with quantum numbers $I(J^P)=0(1^+)$,'' \emph{Phys. Rev. D} \textbf{100}, 014503 (2019), \href{https://arxiv.org/abs/1904.04197}{arXiv:1904.04197}.
\bibitem{ChenReview} H.-X. Chen, W. Chen, X. Liu, and S.-L. Zhu, ``The hidden-charm pentaquark and tetraquark states,'' \emph{Phys. Rep.} \textbf{639}, 1--121 (2016).
\bibitem{GuoReview} F.-K. Guo, C. Hanhart, U.-G. Meißner, Q. Wang, Q. Zhao, and B.-S. Zou, ``Hadronic molecules,'' \emph{Rev. Mod. Phys.} \textbf{90}, 015004 (2018).
\bibitem{BrambillaReview} N. Brambilla \emph{et al.}, ``The XYZ states: experimental and theoretical status and perspectives,'' \emph{Phys. Rep.} \textbf{873}, 1--154 (2020).
\bibitem{LHCbNature} R. Aaij \emph{et al.} (LHCb Collaboration), ``Observation of an exotic narrow doubly charmed tetraquark,'' \emph{Nature Phys.} \textbf{18}, 751--754 (2022), \href{https://doi.org/10.1038/s41567-022-01614-y}{doi:10.1038/s41567-022-01614-y}.
\bibitem{LHCbNC} R. Aaij \emph{et al.} (LHCb Collaboration), ``Study of the doubly charmed tetraquark $T_{cc}^{+}$,'' \emph{Nature Commun.} \textbf{13}, 3351 (2022), \href{https://doi.org/10.1038/s41467-022-30206-w}{doi:10.1038/s41467-022-30206-w}.
\bibitem{Luscher} M. Lüscher, ``Two-particle states on a torus and their relation to the scattering matrix,'' \emph{Nucl. Phys. B} \textbf{354}, 531--578 (1991).
\bibitem{Ortega2023} P. G. Ortega, J. Segovia, D. R. Entem, and F. Fernández, ``Nature of the doubly-charmed tetraquark $T_{cc}^{+}$ in a constituent quark model,'' \emph{Phys. Lett. B} \textbf{841}, 137918 (2023), \href{https://doi.org/10.1016/j.physletb.2023.137918}{doi:10.1016/j.physletb.2023.137918}.
\bibitem{Qiu2024} L. Qiu, C. Gong, and Q. Zhao, ``Coupled-channel description of charmed heavy hadronic molecules within the meson-exchange model and its implication,'' \emph{Phys. Rev. D} \textbf{109}, 076016 (2024), \href{https://doi.org/10.1103/PhysRevD.109.076016}{doi:10.1103/PhysRevD.109.076016}, \href{https://arxiv.org/abs/2311.10067}{arXiv:2311.10067}.
\bibitem{Gil2025} F. Gil-Domínguez, A. Giachino, and R. Molina, ``Quark mass dependence of the $T_{cc}(3875)^+$ pole,'' \emph{Phys. Rev. D} \textbf{111}, 016029 (2025), \href{https://doi.org/10.1103/PhysRevD.111.016029}{doi:10.1103/PhysRevD.111.016029}, \href{https://arxiv.org/abs/2409.15141}{arXiv:2409.15141}.
\bibitem{Prelovsek2025} S. Prelovsek, E. Ortiz-Pacheco, S. Collins, L. Leskovec, M. Padmanath, and I. Vujmilovic, ``Doubly heavy tetraquarks from lattice QCD: Incorporating diquark-antidiquark operators and the left-hand cut,'' \emph{Phys. Rev. D} \textbf{112}, 014507 (2025), \href{https://doi.org/10.1103/rlgp-c9tb}{doi:10.1103/rlgp-c9tb}, \href{https://arxiv.org/abs/2504.03473}{arXiv:2504.03473}.
\bibitem{Whyte2025} T. Whyte, D. J. Wilson, and C. E. Thomas, ``Near-threshold states in coupled $DD^*-D^*D^*$ scattering from lattice QCD,'' \emph{Phys. Rev. D} \textbf{111}, 034511 (2025), \href{https://doi.org/10.1103/PhysRevD.111.034511}{doi:10.1103/PhysRevD.111.034511}, \href{https://arxiv.org/abs/2405.15741}{arXiv:2405.15741}.
\bibitem{Dawid2025} S. M. Dawid, F. Romero-López, and S. R. Sharpe, ``Finite- and infinite-volume study of $DD\pi$ scattering,'' \emph{J. High Energy Phys.} \textbf{01} (2025) 060, \href{https://doi.org/10.1007/JHEP01(2025)060}{doi:10.1007/JHEP01(2025)060}, \href{https://arxiv.org/abs/2409.17059}{arXiv:2409.17059}.
\bibitem{DawidCompare2025} S. M. Dawid, F. Romero-López, and S. R. Sharpe, ``Comparison of integral equations used to study $T_{cc}^{+}$ for a stable $D^*$,'' \emph{J. High Energy Phys.} \textbf{09} (2025) 058, \href{https://doi.org/10.1007/JHEP09(2025)058}{doi:10.1007/JHEP09(2025)058}, \href{https://arxiv.org/abs/2505.05466}{arXiv:2505.05466}.
\bibitem{XuRen2025} H. Xu and L.-X. Ren, ``Investigating the two-pion exchange of the double charm $DD^*$ chiral interactions and $T_{cc}$,'' \emph{Phys. Rev. D} \textbf{112}, 114003 (2025), \href{https://doi.org/10.1103/gryg-xtt4}{doi:10.1103/gryg-xtt4}, \href{https://arxiv.org/abs/2509.11234}{arXiv:2509.11234}.
\bibitem{Kim2024Hidden} H. Kim and K. S. Kim, ``Role of hidden-color components in the tetraquark mixing model,'' \emph{Eur. Phys. J. C} \textbf{84}, 726 (2024), \href{https://doi.org/10.1140/epjc/s10052-024-13072-2}{doi:10.1140/epjc/s10052-024-13072-2}, \href{https://arxiv.org/abs/2403.18411}{arXiv:2403.18411}.
\bibitem{Padmanath2022} M. Padmanath and S. Prelovsek, ``Signature of a doubly charm tetraquark pole in $DD^*$ scattering on the lattice,'' \emph{Phys. Rev. Lett.} \textbf{129}, 032002 (2022), \href{https://arxiv.org/abs/2202.10110}{arXiv:2202.10110}.
\bibitem{Collins2024} S. Collins, A. Nefediev, M. Padmanath, and S. Prelovsek, ``Toward the quark mass dependence of $T_{cc}^{+}$ from lattice QCD,'' \emph{Phys. Rev. D} \textbf{109}, 094509 (2024), \href{https://arxiv.org/abs/2402.14715}{arXiv:2402.14715}.
\bibitem{Du2022} M.-L. Du, V. Baru, X.-K. Dong, A. Filin, F.-K. Guo, C. Hanhart, A. Nefediev, J. Nieves, and Q. Wang, ``Coupled-channel approach to $T_{cc}^{+}$ including three-body effects,'' \emph{Phys. Rev. D} \textbf{105}, 014024 (2022), \href{https://doi.org/10.1103/PhysRevD.105.014024}{doi:10.1103/PhysRevD.105.014024}, \href{https://arxiv.org/abs/2110.13765}{arXiv:2110.13765}.
\bibitem{Lyu2023} Y. Lyu, S. Aoki, T. Doi, T. Hatsuda, Y. Ikeda, and J. Meng, ``Doubly charmed tetraquark $T_{cc}^{+}$ from lattice QCD near physical point,'' \emph{Phys. Rev. Lett.} \textbf{131}, 161901 (2023), \href{https://arxiv.org/abs/2302.04505}{arXiv:2302.04505}.
\bibitem{MonemzadehTazimi2026} M. Monemzadeh and N. Tazimi, ``A Colour-Casimir Adjacency Matrix Approach to Fully-Heavy Tetraquarks,'' arXiv:2609.10822 [hep-ph] (2026), \href{https://arxiv.org/abs/2609.10822}{arXiv:2609.10822}.
\end{thebibliography}
\end{document}